\documentclass{article}%
\usepackage{amsmath}
\usepackage{amsfonts}
\usepackage{amssymb}
\usepackage{graphicx}%
\providecommand{\U}[1]{\protect\rule{.1in}{.1in}}

\begin{document}

\title{Entropy, Information, and the Updating of Probabilities }
\author{Ariel Caticha\\{\small Department of Physics, University at Albany-SUNY, }\\{\small Albany, NY 12222, USA.}}
\date{}
\maketitle

\begin{abstract}
This paper is a review of a particular approach to the method of maximum
entropy as a general framework for inference. The discussion emphasizes the
pragmatic elements in the derivation. An epistemic notion of information is
defined in terms of its relation to the Bayesian beliefs of ideally rational
agents. The method of updating from a prior to a posterior probability
distribution is designed through an eliminative induction process. The
logarithmic relative entropy is singled out as the unique tool for updating
that (a)\ is of universal applicability; (b) that recognizes the value of
prior information; and (c) that recognizes the privileged role played by the
notion of independence in science.

The resulting framework --- the ME\ method --- can handle arbitrary priors and
arbitrary constraints. It includes MaxEnt and Bayes' rule as special cases
and, therefore, it unifies entropic and Bayesian methods into a single general
inference scheme. The ME method goes beyond the mere selection of a single
posterior, but also addresses the question of how much less probable other
distributions might be, which provides a direct bridge to the theories of
fluctuations and large deviations.

\end{abstract}

\section{Introduction}

Inductive inference is a framework for coping with uncertainty, for reasoning
with incomplete information. The framework must include a means to represent a
state of partial knowledge --- this is handled through the introduction of
probabilities --- and it must allow us to change from one state of partial
knowledge to another when new information becomes available. Indeed any
inductive method that recognizes that a situation of incomplete information is
in some way unfortunate --- by which we mean that it constitutes a problem in
need of a solution --- would be severely deficient if it failed to address the
question of how to proceed in those fortunate circumstances when new
information becomes available. The theory of probability, if it is to be
useful at all, demands a method for assigning and updating probabilities.

The challenge is to develop updating methods that are both systematic,
objective and practical. When the information consists of data and a
likelihood function Bayesian updating is the uniquely natural method of
choice. Its foundation lies on recognizing the value of prior information:
whatever was learned in the past is valuable and should not be disregarded,
which amounts to requiring that beliefs ought to be revised but only to the
extent required by the new data.\emph{ }This immediately raises a number of
questions: How do we update when the information is not in the form of data?
If the information is not data, what else could it possibly be? Indeed what,
after all, is `information'? On a separate line of development the method of
Maximum Gibbs-Shannon Entropy (MaxEnt) allows one to process information in
the form of constraints on the allowed probability distributions. This
provides a partial answer to one of our questions: in addition to data,
information can also take the form of constraints. But it immediately raises
several other questions: What is the interpretation of entropy? Is there a
unique entropy? Are Bayesian and entropic methods mutually compatible?

The purpose of this paper is to review one particular approach to entropic
updating. The presentation below, which is meant to be pedagogical andOn a
separate line of development the method of Maximum Gibbs-Shannon Entropy
(MaxEnt) allows one to process information inOn a separate line of development
the method of Maximum Gibbs-Shannon Entropy (MaxEnt) allows one to process
information in self-contained, is based on work presented in a sequence of
papers \cite{Caticha 2003}\cite{Caticha Giffin 2006}\cite{Caticha
2007}\cite{Caticha 2014}\cite{Vanslette 2017} and in the sets of lectures
\cite{Caticha 2008}\cite{Caticha 2012}\cite{Caticha 2020}. As we shall see
below we adopt a pragmatic approach in which entropy is a tool designed for
the specific purpose of updating probabilities.

Historically the method of Maximum relative Entropy (ME) is a direct
descendant of the MaxEnt method pioneered by Jaynes \cite{Jaynes
1957}\cite{Jaynes 1983}. In the MaxEnt framework entropy is interpreted
through the Shannon axioms as a measure of the amount of information\ that is
missing\ in a probability distribution. This approach has its limitations. The
Shannon axioms refer to probabilities of discrete variables; for continuous
variables the Shannon entropy is not defined. A more serious objection is that
even if we grant that the Shannon axioms do lead to a reasonable expression
for the entropy, to what extent do we believe the axioms themselves? Shannon's
third axiom, the grouping property, is indeed sort of reasonable, but is it
necessary? Is entropy the only consistent measure of uncertainty or of
information? Indeed, there exist examples in which the Shannon entropy does
not seem to reflect one's intuitive notion of information \cite{Caticha
2020}\cite{Uffink 1995}. One could introduce other entropies justified by
different choices of axioms (\emph{e.g.}, \cite{Renyi 1961}\cite{Aczel Daroczy
1975}\cite{Tsallis 1988}). But this move raises problems of its own: Which
entropy should one adopt? If different systems are to handled using different
entropies, how does one handle composite systems?

From our perspective the problem can be traced to the fact that neither
Shannon nor Jaynes were concerned with the task of updating probabilities.
Shannon's communication theory aimed to characterize the sources of
information, to measure the capacity of the communication channels, and to
learn how to control the degrading effects of noise. On the other hand, Jaynes
conceived MaxEnt as a method to assign probabilities on the basis of
constraint information and a fixed underlying measure and not from an
arbitrary prior.

Considerations such as these motivated several attempts to develop ME directly
as a method for updating probabilities without invoking questionable measures
of information \cite{Caticha 2003}\cite{Vanslette 2017}\cite{Shore Johnson
1980}\cite{Skilling 1988}\cite{Skilling 1989}. The important contribution by
Shore and Johnson was the realization that one could axiomatize the updating
method itself rather than the information measure. Their axioms have, however,
raised criticisms \cite{Uffink 1995}\cite{Karbelkar 1986}\cite{Tsallis
2015}\cite{Jizba Korbel 2019}\ and counter-criticisms \cite{Caticha Giffin
2006}\cite{Caticha 2008}\cite{Caticha 2020}\cite{Presse et al 2013}%
\cite{Presse et al 2015}. Despite the controversies, Shore and Johnson's
pioneering papers have had an enormous influence: they identified the correct
goal to be achieved.

The concept of relative entropy is introduced as a tool for updating
probabilities. From now we drop the qualifier \textquotedblleft
relative\textquotedblright and adopt the simpler term \textquotedblleft
entropy\textquotedblright. The reasons for the improved nomenclature are the
following: (1) the general concept should receive the general name
\textquotedblleft entropy\textquotedblright, while the more specialized
concepts should be the ones receiving a qualifier such as \textquotedblleft
thermodynamic\textquotedblright\ or \textquotedblleft
Clausius\textquotedblright\ entropy, and \textquotedblleft
Gibbs-Shannon\textquotedblright\ entropy. (2) All entropies are relative, even
if they happen to be relative to an implicit uniform prior. Making this fact
explicit has tremendous pedagogical value. (3) The practice is already in use
with the concept of energy: all energies are relative too, but there is no
advantage in constantly referring to a \textquotedblleft relative
energy\textquotedblright. Accordingly, ME will be read as \textquotedblleft
Maximum Entropy\textquotedblright; additional qualifiers are redundant.

As with all tools, entropy too is \emph{designed} to perform a certain
function and its performance must meet certain \emph{design criteria
}or\emph{\ specifications}. There is no implication that the method is
\textquotedblleft true\textquotedblright, or that it succeeds\ because it
achieves some special contact with reality. Instead the claim is that the
method succeeds in the pragmatic sense that it works as designed --- and that
this is satisfactory because when properly deployed it leads to empirically
adequate models. In this approach \emph{entropy needs no interpretation}
whether it be in terms of heat, multiplicity of states, disorder, uncertainty,
or even in terms of an amount of information. \noindent Incidentally, this may
explain why the search for the meaning of entropy has proved so elusive: we
need not know what `entropy' means; we only need to know how to use it.

Since our topic is the updating of probabilities when confronted with new
information, our starting point is to address the question `what is
information?'. In Section 2 we develop a concept of information that is both
pragmatic and Bayesian. `Information' is defined in terms of its effects on
the beliefs of rational agents. The design of entropy as a tool for updating
is the topic of Section 3. There we state the design specifications that
define what function entropy is supposed to perform and we derive its
functional form. To streamline the presentation some of the mathematical
derivations are left to the appendices.

To conclude we present two further developments. In Section 4 we show that
Bayes' rule can be derived as a special case of the ME method. An earlier
derivation of this important result following a different line of argument was
given by Williams \cite{Williams 1980} before a sufficient understanding of
entropy as an updating tool had been achieved. It is not, therefore,
surprising that Williams' achievement has not received the widespread
appreciation it deserves. Thus, within the ME framework entropic and Bayesian
methods are unified into a single consistent theory of inference. One
advantage of this insight is that it allows a number of generalizations of
Bayes' rule \cite{Caticha Giffin 2006}\cite{Caticha 2020}. Another is that it
provides an important missing piece for the old puzzles of quantum mechanics
concerning the so-called collapse of the wave function and the quantum
measurement problem \cite{Johnson Caticha 2011}\cite{Vanslette Caticha 2017}.

There is yet another function that the ME method must perform in order to
fully qualify as a method of inductive inference. Once we have decided that
the distribution of maximum entropy is to be preferred over all others the
following question arises immediately: the maximum of the entropy functional
is never infinitely sharp, are we really confident that distributions that lie
very close to the maximum are totally ruled out? In Section 5 the ME method is
deployed to assess quantitatively the extent to which distributions with lower
entropy are ruled out. The significance of this result is that it provides a
direct link to the theories of fluctuations and large deviations. Concluding
remarks are given in Section 6.

\section{What is information?}

The term `information' is used with a wide variety of different meanings
\cite{Jaynes 1983}\cite{Cover Thomas 1991}\cite{Golan 2018}. There is the
Shannon notion of information that is meant to measure an amount of
information and is quite divorced from semantics. There is also an algorithmic
notion of information that captures a notion of complexity and originates in
the work of Solomonov, Kolmogorov, and Chaitin \cite{Cover Thomas 1991}, and
there is a related notion of entropy as a Minimum Description Length
\cite{Rissanen 1978}. Furthermore, in the general context of the
thermodynamics of computation it is said that \textquotedblleft information is
physical\textquotedblright\ because systems \textquotedblleft
carry\textquotedblright\ or \textquotedblleft contain\textquotedblright%
\ information about their own physical state\ \cite{Landauer 1991}%
\cite{Bennett 1982}\cite{Bennett 2003} (see also \cite{Norton 2011}%
\cite{Norton 2013}).

Here we follow a different path \cite{Caticha 2007}\cite{Caticha 2014}. We
seek an epistemic notion of information that is closer to the everyday
colloquial use of the term --- roughly, information is what we request when we
ask a question. In a Bayesian framework this requires an explicit account of
the relation between information and the beliefs of ideally rational agents
\cite{footnote 1}.

It is implicit in the recognition that most of our beliefs are held on the
basis of incomplete information that not all probability assignments are
equally good; some beliefs are preferable to others in the very pragmatic
sense that they enhance our chances to successfully navigate this world. Thus
a theory of probability demands a theory of updating probabilities in order to
improve our beliefs.

We are now ready to address the question: What, after all, is `information'?
The answer is pragmatic. \emph{Information is what information does.}
Information is defined by its effects: (a) it restricts our options as to what
we are honestly and rationally allowed to believe; and (b) it induces us to
update from prior beliefs to posterior beliefs. This, I propose, is a defining
characteristic of information:\medskip

\noindent\emph{Information is that which induces a change from one state of
rational belief to another.}\medskip

\noindent One aspect of this notion is that for a rational agent, the
identification of what constitutes information --- as opposed to mere noise
--- already involves a judgement, an evaluation. Another aspect is that the
notion that information is directly related to changing our minds does not
involve any talk about \emph{amounts} of information but, it nevertheless
allows precise quantitative calculations. Indeed, constraints on the
acceptable posterior probabilities are precisely the kind of information the
method of maximum entropy is designed to handle. In short,\medskip

\noindent\emph{Information constrains probability distributions. The
constraints are the information.\medskip}

\noindent To the extent that the probabilities are Bayesian, this definition
captures the Bayesian notion that information is directly related to changing
our minds; that it is the driving force behind the process of learning. It
also incorporates an important feature of rationality: being rational means
accepting that \textquotedblleft not everything goes\textquotedblright, that
our beliefs must be constrained in very specific ways. But the indiscriminate
acceptance of any arbitrary constraint does not qualify as rational behavior.
To be rational an agent must exercise some judgement before accepting a
particular piece of information as a reliable basis for the revision of its
beliefs which raises questions about what judgements might be considered
sound. Furthermore, there is no implication that the information must \emph{be
true}; only that we \emph{accept it as true}. False information is information
too, at least to the extent that we are prepared to accept it and allow it to
affect our beliefs.

The paramount virtue of the definition above is that it is useful; it allows
precise quantitative calculations. The constraints that constitute information
can take a wide variety of forms. They can be expressed in terms of expected
values, or they can specify the functional form of a distribution, or be
imposed through various geometrical relations. Examples are given in Section 5
and in \cite{Pessoa et al 2021}.

Concerning the act of updating it may be worthwhile to point out an analogy
with dynamics. In Newtonian mechanics the state of motion of a system is
described in terms of momentum and the change from one state to another is
said to be \textquotedblleft caused\textquotedblright\ by an applied force or
impulse. Bayesian inference is analogous in that a state of belief is
described in terms of probabilities and the change from one state to another
is \textquotedblleft caused\textquotedblright\ by information. Just as a force
is that which induces a change from one state of motion to another, so
\emph{information is that which induces a change from one state of belief to
another}. Updating is a form of dynamics. In \cite{Caticha 2019} the analogy
is taken seriously: the logic is reversed and quantum mechanics is derived as
an example of the entropic updating of probabilities.

\section{The pragmatic design of entropic inference}

Once we have decided, as a result of the confrontation of new information with
old beliefs, that our beliefs require revision the problem becomes one of
deciding how precisely this ought to be done. First we identify some general
features of the kind of belief revision that one might count as rational. Then
we design a method, a systematic procedure, that implements those features. To
the extent that the method performs as desired we can claim success. The point
is not that success derives from our method having achieved some intimate
connection to the inner wheels of reality; success just means that the method
seems to be working.

The one obvious requirement is that the updated probabilities ought to agree
with the newly acquired information. Unfortunately, this requirement, while
necessary, is not sufficiently restrictive: we can update in many ways that
preserve both internal consistency and consistency with the new information.
Additional criteria are needed. What rules would an ideally rational agent choose?

\subsection{General criteria\ }

The rules are motivated by the same pragmatic design criteria that motivate
the design of probability theory itself \cite{Caticha 2020} --- universality,
consistency, and practical utility. But this is admittedly too vague; we must
be very specific about the precise way in which they are implemented.

\subsubsection{Universality}

In principle different systems and different situations could require
different problem-specific induction methods. However, in order to be useful
in practice, the method we seek must be of \emph{universal} applicability.
Otherwise it would fail us when most needed, for we would not know which
method to choose when not much is known about the system. To put in different
words, what we want to design is a general-purpose method that captures what
all the other problem-specific methods might have in common. The idea is that
the peculiarities of a particular problem will be captured by the specific
constraints that describe the information that is relevant to the problem at hand.

The analogy with mechanics can be found here too. The possibility of a science
of mechanics hinges on identifying a law of motion of universal applicability
(\emph{e.g.}, the Schr\"{o}dinger equation) while the specifics of each system
are introduced through initial conditions and the choice of potentials or
forces. Here we shall design an entropy of universal applicability, while the
specifics of each problem are introduced through prior probabilities and the
choice of constraints.

\subsubsection{Parsimony}

To specify the updating we adopt a very conservative criterion that recognizes
the value of information: what has been laboriously learned in the past is
valuable and should not be disregarded unless rendered obsolete by new
information. The only aspects of one's beliefs that should be updated are
those for which new evidence has been supplied. Thus we adopt a\medskip

\noindent\textbf{Principle of Minimal Updating (PMU)}: \emph{Beliefs should be
updated only to the minimal extent required by the new information.}\medskip

\noindent The special case of updating in the absence of new information
deserves a comment. The PMU states that when there is no new information
ideally rational agents should not change their minds. In fact, it is
difficult to imagine any notion of rationality that would allow the
possibility of changing one's mind for no apparent reason.

Minimal updating offers yet another pragmatic advantage. As we shall see
below, rather than identifying what features of a distribution are singled out
for updating and then specifying the detailed nature of the update, we will
adopt design criteria that stipulate what is not to be updated. The practical
advantage of this approach is that it enhances objectivity --- there are many
ways to change something but only one way to keep it the same. The analogy
with mechanics can be pursued even further: if updating is a form of dynamics,
then minimal updating is the analogue of inertia. Rationality and objectivity
demand a considerable amount of inertia.

\subsubsection{Independence}

The next general requirement turns out to be crucially important because
without it the very possibility of scientific theories would be compromised.
The point is that every scientific model, whatever the topic, if it is to be
useful at all, must assume that all relevant variables have been taken into
account and that whatever was left out --- the rest of the universe --- should
not matter. To put it another way, in order to do science we must be able to
understand parts of the universe without having to understand the universe as
a whole. Granted, a pragmatic understanding need not be complete and exact; it
must be merely adequate for our purposes.

The assumption, then, is that it is possible to focus our attention on a
suitably chosen system of interest and neglect the rest of the universe
because the system and the rest of the universe are \textquotedblleft
sufficiently independent\textquotedblright.\ Thus, in any form of science the
notion of statistical independence must play a central and privileged role.
This idea --- that some things can be neglected, that not everything matters
--- is implemented by imposing a criterion that tells us how to handle
independent systems. The chosen criterion is quite natural: \emph{Whenever two
systems are a priori believed to be independent and we receive information
about just one, it should not matter if the other is included in the analysis
or not.} This is an example of the PMI in action; it amounts to requiring that
independence be preserved unless information about correlations is explicitly introduced.

Again we emphasize:\ none of these criteria are imposed by Nature. They are
desirable for pragmatic reasons; they are imposed by design.

\subsection{Entropy as a tool for updating probabilities}

Consider a set of propositions $\{x\}$ about which we are uncertain. The
proposition $x$ can be discrete or continuous, in one or in several
dimensions. It could, for example, represent the microstate of a physical
system, a point in phase space, or an appropriate set of quantum numbers. The
uncertainty about $x$ is described by a probability distribution $q(x)$. The
goal is to update from the prior distribution $q(x)$ to a posterior
distribution $p(x)$ when new information --- by which we mean a set of
constraints --- becomes available. The question is: which distribution among
all those that satisfy the constraints should we select?

Our goal is to design a method that allows a systematic search for the
preferred posterior distribution. The central idea, first proposed by Skilling
\cite{Skilling 1988}, is disarmingly simple: to select the posterior first
rank all candidate distributions in increasing order of \textquotedblleft
preference\textquotedblright\ and then pick the distribution that ranks the
highest. Irrespective of what it is that makes one distribution
\textquotedblleft preferable\textquotedblright\ over another (we will get to
that soon enough) it is clear that any such ranking must be transitive: if
distribution $p_{1}$ is preferred over distribution $p_{2}$, and $p_{2}$ is
preferred over $p_{3}$, then $p_{1}$ is preferred over $p_{3}$. Transitive
rankings are implemented by assigning to each $p$ a real number $S[p]$, which
is called the entropy of $p$, in such a way that if $p_{1}$ is preferred over
$p_{2}$, then $S[p_{1}]>S[p_{2}]$. The selected distribution (one or possibly
many, for \emph{there may be several equally preferred distributions}) is that
which maximizes the entropy functional.

The importance of Skilling's strategy of ranking distributions cannot be
overestimated: it answers the questions `why an entropy?' and `why a
maximum?'. The strategy implies that the updating method will take the form of
a variational principle --- the method of Maximum Entropy (ME) --- and that
the latter will involve a certain functional that maps distributions to real
numbers and that is designed to be maximized. These features are not imposed
by Nature; they are all imposed by design. They are dictated by the function
that the ME method is supposed to perform. (Thus, it makes no sense to seek a
generalization in which entropy is a complex number or a vector; such
generalized entropies would just not perform the desired function.)

Next we specify the ranking scheme, that is, we choose a specific functional
form for the entropy $S[p]$. Note that the purpose of the method is to
update\emph{ from priors to posteriors} so the ranking scheme must depend on
the particular prior $q$ and therefore the entropy $S$ must be a functional of
both $p$ and $q$. The entropy $S[p,q]$ describes a ranking of the
distributions $p$ \emph{relative} to the given prior $q$. $S[p,q]$ is the
entropy of $p$ \emph{relative} to $q$, and accordingly $S[p,q]$ is commonly
called \emph{relative entropy}. This is appropriate and sometimes we will
follow this practice. However, since all entropies are relative, even when
relative to a uniform distribution, the qualifier `relative' is redundant and
can be dropped.

The functional $S[p,q]$ is designed by a process of elimination --- this is a
process of \emph{eliminative induction}. First we state the desired design
criteria; this is the crucial step that defines what makes one distribution
preferable over another. Candidate functionals that fail to satisfy the
criteria are discarded --- hence the qualifier `eliminative'. As we shall see
the criteria adopted below are sufficiently constraining that there is a
single entropy functional $S[p,q]$ that survives the process of elimination.

This approach has a number of virtues. First, to the extent that the design
criteria are universally desirable, the single surviving entropy functional
will be of universal applicability too. Second, the reason why alternative
entropy candidates are eliminated is quite explicit --- at least one of the
design criteria is violated. Thus, \emph{the justification behind the single
surviving entropy is not that it leads to demonstrably correct inferences, but
rather, that all other candidates demonstrably fail to perform as desired.}

\subsection{Specific design criteria}

\noindent\label{specific DC}Consider a lattice of propositions generated by a
set $\mathcal{X}$ of atomic propositions that are mutually exclusive and
exhaustive and are labeled by a discrete index $i=1,2\ldots n$. The extension
to infinite sets and to continuous labels turns out to be straightforward. The
index $i$ might, for example, label the microstates of a physical system but,
since the argument below is supposed to be of general validity, we shall not
assume that the labels themselves carry any particular significance. We can
always permute labels and this should have no effect on the updating of probabilities.

We adopt design criteria that reflect the structure of the lattice of
propositions --- the propositions are related to each other by disjunctions
(\textsc{or}) and conjunctions (\textsc{and}) and the consistency of the web
of beliefs is implemented through the sum and product rules of probability
theory. Our criteria refer to the two extreme situations of propositions that
are mutually exclusive and of propositions that are mutually independent. At
one end we deal with the probabilities of propositions that are highly
correlated (if one proposition is true the other is false and vice versa); at
the other end we deal with the probabilities of propositions that are totally
uncorrelated (the truth or falsity of one proposition has no effect on the
truth or falsity of the other). One extreme is described by the simplified sum
rule, $p(i\vee j)=p(i)+p(j)$, and the other extreme by the simplified product
rule, $p(i\wedge j)=p(i)p(j)$ \cite{footnote 2}.

The two design criteria and their consequences for the functional form of the
entropy are given below. Detailed proofs are deferred to the appendices.

\subsubsection{Mutually exclusive subdomains}

\label{DC1}

\noindent\textbf{DC1}: \emph{Probabilities that are conditioned on one
subdomain are not affected by information about other non-overlapping
subdomains.\medskip}

\noindent Consider a subdomain $\mathcal{D\subset X}$ composed of atomic
propositions $i\in\mathcal{D}$ and suppose the information to be processed
refers to some other subdomain $\mathcal{D}^{\prime}\mathcal{\subset X}$ that
does not overlap with $\mathcal{D}$, $\mathcal{D\cap D}^{\prime}=\emptyset$.
In the absence of any new information about $\mathcal{D}$ the PMU demands we
do not change our minds about probabilities that are conditional on
$\mathcal{D}$. Thus, we design the inference method so that $q(i|\mathcal{D}%
)$, the prior probability of $i$ conditioned on $i\in\mathcal{D}$, is not
updated. Thus, the selected conditional posterior is
\begin{equation}
P(i|\mathcal{D)}=q(i|\mathcal{D})~. \label{DC1a}%
\end{equation}
(We adopt the following notation: priors are denoted by $q$, candidate
posteriors by lower case $p$, and the selected posterior by upper case $P$. We
shall write either $p(i)$ or $p_{i}$. Furthermore, we adopt the notation,
standard in physics where the probabilities of $x$ and $\theta$ are written
$p(x)$ and $p(\theta)$ but there is no implication that $p$ refers to a single
mathematical function.)

We emphasize: the point is not that we make the unwarranted assumption that
keeping $q(i|\mathcal{D})$ unchanged is guaranteed to lead to correct
inferences. It need not; induction is risky. The point is, rather, that in the
absence of any evidence to the contrary there is no reason to change our minds
and the prior information takes priority.

\noindent\textbf{The consequence of DC1} is that non-overlapping domains of
$i$ contribute additively to the entropy,
\begin{equation}
S(p,q)=%
{\displaystyle\sum\limits_{i}}
\,F\left(  p_{i},q_{i}\right)  \ , \label{DC1b}%
\end{equation}
where $F$ is some unknown function of two arguments. The proof is given in
Appendix A.

\noindent\textbf{Comment 1: }It is essential that DC1 refers to
\emph{conditional} probabilities: local information about a domain
$\mathcal{D}^{\prime}$ can have a non-local effect on the total probability of
another domain $\mathcal{D}$.

\noindent\textbf{Comment 2: }An important special case is the
\textquotedblleft update\textquotedblright\ from a prior $q(i)$ to a posterior
$P(i)$ in a situation in which no new information is available. The criterion
DC1 applied to a situation where the subdomain $\mathcal{D}$ covers the whole
space of $i$s, $\mathcal{D=X}$, requires that \emph{in the absence of any new
information the prior conditional probabilities are not to be updated:
}$P(i|\mathcal{X})=q(i|\mathcal{X})$\emph{\ or }$P(i)=q(i)$.

\noindent\textbf{Comment 3: }The criterion DC1 implies Bayesian
conditionalization as a special case. Indeed, if the information is given
through the constraint $p(\mathcal{\tilde{D}})=0$ where $\mathcal{\tilde{D}}$
is the complement of $\mathcal{D}$ then $P(i|\mathcal{D})=q(i|\mathcal{D})$,
which is referred to as Bayesian conditionalization. More explicitly, if
$\theta$ is the variable to be inferred on the basis of prior information
about a likelihood function $q(i|\theta)$ and observed data $i^{\prime}$, then
the update from the prior $q$ to the posterior $P$,
\begin{equation}
q(i,\theta)=q(i)q(\theta|i)\rightarrow P(i,\theta)=P(i)P(\theta|i)
\end{equation}
consists of updating $q(i)\rightarrow P(i)=\delta_{ii^{\prime}}$ to agree with
the new information and invoking the PMU so that $P(\theta|i^{\prime
})=q(\theta|i^{\prime})$ remains unchanged. Therefore,
\begin{equation}
P(i,\theta)=\delta_{ii^{\prime}}q(\theta|i^{\prime})\quad\text{so that}\quad
P(\theta)=q(\theta|i^{\prime})=q(\theta)\frac{q(i^{\prime}|\theta
)}{q(i^{\prime})}~,
\end{equation}
which is Bayes' rule. Thus, \emph{entropic inference is designed to include
Bayesian inference as a special case}. Note however that imposing DC1 is not
identical to imposing Bayesian conditionalization: DC1 is not restricted to
information in the form of absolute certainties such as $p(\mathcal{D})=1$.

\noindent\textbf{Comment 4: }If the label $i$ is turned into a continuous
variable $x$ the criterion DC1 requires that information that refers to points
infinitely close but just outside the domain $\mathcal{D}$ will have no
influence on probabilities conditional on $\mathcal{D}$. This may seem
surprising as it may lead to updated probability distributions that are
discontinuous, but it is not a problem. In situations where we have explicit
reasons to believe that conditions of continuity or differentiability hold,
then such conditions should be imposed explicitly. The inference process
should not be expected to discover and replicate information with which it was
not supplied.

\subsubsection{Subsystem independence}

\label{DC2}

\noindent\textbf{DC2}: \emph{When two systems are a priori believed to be
independent and the information we receive about one of them makes no
reference to the other, then it should not matter whether the latter is
included in the analysis of the former or not.\medskip}

Consider a system of propositions labelled by a composite index,
$i=(i_{1},i_{2})\in\mathcal{X}=\mathcal{X}_{1}\times\mathcal{X}_{2}$. For
example, $\{i_{1}\}=\mathcal{X}_{1}$ and $\{i_{2}\}=\mathcal{X}_{2}$ might
describe the microstates of two separate physical systems. Assume that all
prior evidence led us to believe the two subsystems are independent, that is,
any two propositions $i_{1}\in\mathcal{X}_{1}$ and $i_{2}\in\mathcal{X}_{2}$
are believed to be independent. This belief is reflected in the prior
distribution: if the individual subsystem priors $q_{1}(i_{1})$ and
$q_{2}(i_{2})$, then the prior for the whole system is $q_{1}(i_{1}%
)q_{2}(i_{2})$. Next suppose that new information is acquired such that
$q_{1}(i_{1})$ would by itself be updated to $P_{1}(i_{1})$, and that
$q_{2}(i_{2})$ would by itself be updated to $P_{2}(i_{2})$. DC2 requires that
$S[p,q]$ be such that the joint prior $q_{1}(i_{1})q_{2}(i_{2})$ updates to
the product $P_{1}(i_{1})P_{2}(i_{2})$ so that inferences about one subsystem
do not affect inferences about the other.

\noindent\textbf{The consequence of DC2 }is to fully determine the unknown
function $F$ in (\ref{DC1b}) so that probability distributions $p(i)$ should
be ranked relative to the prior $q(i)$ according to the relative entropy,
\begin{equation}
S[p,q]=-%
{\displaystyle\sum\limits_{i}}
\,p(i)\log\frac{p(i)}{q(i)}. \label{S[p,q] a}%
\end{equation}

\noindent\textbf{Comment 1: }We emphasize that the point is not that when we
have no evidence for correlations we draw the firm conclusion that the systems
must necessarily be independent. Induction involves risk; the systems might in
actual fact be correlated through some unknown interaction potential. The
point is rather that if the joint prior reflected independence and the new
evidence is silent on the matter of correlations, then the evidence we
actually have --- namely, the prior --- takes precedence and there is no
reason to change our minds. As before, the PMU requires that a feature of the
probability distribution --- in this case, independence --- will not be
updated unless the evidence requires it.

\noindent\textbf{Comment 2: }We also emphasize that DC2\emph{ is not a
consistency requirement}. The argument we deploy is \emph{not} that both the
prior \emph{and} the new information tell us the systems are independent in
which case consistency requires that it should not matter whether the systems
are treated jointly or separately. DC2 refers to a situation where the new
information does not say whether the systems are independent or not. Rather,
the updating is being \emph{designed} --- through the PMU --- so that the
independence reflected in the prior is maintained in the posterior by default.

\noindent\textbf{Comment 3: }The generalization to continuous variables
$x\in\mathcal{X}$ is approached as a Riemann limit from the discrete case. A
continuous probability density $p(x)$ or $q(x)$ can be approximated by the
discrete distributions. Divide the region of interest $\mathcal{X}$ into a
large number $N$ of small cells. The probabilities of each cell are
\begin{equation}
p_{i}=p(x_{i})\Delta x_{i}\quad\text{and}\quad q_{i}=q(x_{i})\Delta x_{i}~,
\end{equation}
where $\Delta x_{i}$ is an appropriately small interval. The discrete entropy
of $p_{i}$ relative to $q_{i}$ is%
\begin{equation}
S_{N}=-\sum_{i=1}^{N}\Delta x_{i}\,p(x_{i})\,\log\left[  \frac{p(x_{i})\Delta
x_{i}}{q(x_{i})\Delta x_{i}}\right]  \,~,
\end{equation}
and in the limit as $N\rightarrow\infty$ and $\Delta x_{i}\rightarrow0$ we get
the Riemann integral
\begin{equation}
S[p,q]=-\int dx\,p(x)\,\log\,\left[  \frac{p(x)}{q(x)}\right]  ~,
\label{S[p,q] b}%
\end{equation}
(To simplify the notation we include multi-dimensional integrals by writing
$d^{n}x=dx$.) It is easy to check that the ranking of distributions induced by
$S[p,q]$ is invariant under coordinate transformations. The insight that
coordinate invariance could be derived as a consequence of the requirement of
subsystem independence first appeared in \cite{Vanslette 2017}.

\subsection{The ME\ method}

\label{The ME method}

We can now summarize the overall conclusion:\medskip

\noindent\textbf{The ME method}: \emph{The goal is to update from a prior
distribution }$q$\emph{\ to a posterior distribution\ when there is new
information in the form of constraints }$\mathcal{C}$\emph{\ that specify a
family }$\{p\}$\emph{\ of candidate posteriors. The preferred posterior }$P$
\emph{is that which maximizes the relative entropy, }%
\begin{equation}
S[p,q]=-%
{\displaystyle\sum\limits_{i}}
\,p_{i}\log\frac{p_{i}}{q_{i}}\quad\text{\emph{or}}\quad S[p,q]=-\int
dx\,p(x)\,\log\,\left[  \frac{p(x)}{q(x)}\right]  ~, \label{S[p,q]}%
\end{equation}
\emph{\ within the family }$\{p\}$\emph{ specified by the constraints
}$\mathcal{C}$\emph{.}\medskip

This extends the method of maximum entropy beyond its original purpose as a
rule to assign probabilities from a given underlying measure (MaxEnt) to a
method for updating probabilities from any arbitrary prior (ME). Furthermore,
the logic behind the updating procedure does not rely on any particular
meaning assigned to the entropy, either in terms of information, or heat, or
disorder. Entropy is merely a tool for inductive inference. \emph{No
interpretation for }$S[p,q]$\emph{\ is given and none is needed. }

The derivation above has singled out \emph{a unique }$S[p,q]$\emph{\ to be
used in inductive inference}. Other \textquotedblleft
entropies\textquotedblright\ (such as, the one-parameter families of entropies
proposed in \cite{Renyi 1961}\cite{Tsallis 1988}\cite{Aczel Daroczy 1975}
might turn out to be useful for other purposes --- perhaps as measures of some
kind of \textquotedblleft information\textquotedblright, or as measures of
discrimination or distinguishability among distributions, or of ecological
diversity, or for some altogether different function --- but they are
unsatisfactory for the purpose of updating because they fail to perform the
functions stipulated by the design criteria DC1 and DC2. They induce
correlations that are unwarranted by the information in the priors or the constraints.

\section{Bayes' rule as a special case of ME}

Back in subsection \ref{DC1} we saw that ME is designed to include Bayes' rule
as a special case. Here we wish to verify this explicitly \cite{Caticha Giffin
2006}. The goal is to update our beliefs about $\theta\in\Theta$ ($\theta$
represents one or many parameters) on the basis of three pieces of
information: (1) the prior information codified into a prior distribution
$q(\theta)$; (2) the new information conveyed by data $x\in\mathcal{X}$
(obtained in one or many experiments); and (3) the known relation between
$\theta$ and $x$ given by a model defined by the sampling distribution or
likelihood, $q(x|\theta)$. The updating will result in replacing the
\emph{prior} probability distribution $q(\theta)$ by a \emph{posterior}
distribution $P(\theta)$ that applies after the data information has been processed.

The crucial element that will allow Bayes' rule to be smoothly integrated into
the ME scheme is the realization that before the data is collected not only we
do not know $\theta$, we do not know $x$ either. Thus, the relevant space for
inference is not the space $\Theta$ but the product space $\Theta
\times\mathcal{X}$ and the relevant joint \emph{prior} is $q(x,\theta
)=q(\theta)q(x|\theta)$. Let us emphasize two points: first, the likelihood
function is an integral part of the \emph{prior} distribution; and second, the
prior information about how $x$ is related to $\theta$ is contained in the
\emph{functional form} of the distribution $q(x|\theta)$ and not in the
numerical values of the arguments $x$ and $\theta$ which, at this point, are
still unknown.

Next, data is collected and the observed values turn out to be $x^{\prime}$.
We must update to a posterior that lies within the family of distributions
$p(x,\theta)$ that reflect the fact that the previously unknown $x$ is now
known to be $x^{\prime}$, that is,
\begin{equation}
p(x)=\int d\theta\,p(\theta,x)=\delta(x-x^{\prime})~.
\label{data constraint a}%
\end{equation}
The information in this data constrains but is not sufficient to fully
determine the joint distribution
\begin{equation}
p(x,\theta)=p(x)p(\theta|x)=\delta(x-x^{\prime})p(\theta|x^{\prime})~.
\end{equation}
Any choice of $p(\theta|x^{\prime})$ is in principle possible. So far the
formulation of the problem parallels subsection \ref{DC1} exactly. We are,
after all, solving the same problem. The next step is to apply the ME method.

According to the ME method the selected joint posterior $P(x,\theta)$ is that
which maximizes the entropy,%
\begin{equation}
S[p,q]=-\int dxd\theta~p(x,\theta)\log\frac{p(x,\theta)}{q(x,\theta)}~,~
\label{entropy}%
\end{equation}
subject to the data constraints. Note that eq.(\ref{data constraint a})
represents an \emph{infinite} number of constraints on the family
$p(x,\theta)$: there is one constraint and one Lagrange multiplier
$\lambda(x)$ for each value of $x$. Maximizing $S$, (\ref{entropy}), subject
to (\ref{data constraint a}) and normalization,
\begin{align}
\delta\left\{  S+\alpha\left[
{\textstyle\int}
dxd\theta~p(x,\theta)-1\right]  +%
{\textstyle\int}
dx\,\lambda(x)\left[
{\textstyle\int}
d\theta~p(x,\theta)-\delta(x-x^{\prime})\right]  \right\}   &  =0~,\nonumber\\
&  {}%
\end{align}
yields the joint posterior,
\begin{equation}
P(x,\theta)=q(x,\theta)\,\frac{e^{\lambda(x)}}{Z}~,
\end{equation}
where $Z$ is a normalization constant, and the multiplier $\lambda(x)$ is
determined from (\ref{data constraint a}),
\begin{equation}%
{\textstyle\int}
d\theta~q(x,\theta)\frac{\,e^{\lambda(x)}}{Z}=q(x)\frac{\,e^{\lambda(x)}}%
{Z}=\delta(x-x^{\prime})~,
\end{equation}
so that the joint posterior is%
\begin{equation}
P(x,\theta)=q(x,\theta)\frac{\,\delta(x-x^{\prime})}{q(x)}=\delta(x-x^{\prime
})q(\theta|x)~.
\end{equation}
The corresponding marginal posterior probability $P(\theta)$ is
\begin{equation}
P(\theta)=%
{\textstyle\int}
dx\,P(\theta,x)=q(\theta|x^{\prime})=q(\theta)\frac{q(x^{\prime}|\theta
)}{q(x^{\prime})}~, \label{Bayes rule}%
\end{equation}
which is Bayes' rule. Thus, Bayes' rule is derivable from and therefore
consistent with the ME method.

To summarize: the prior $q(x,\theta)=q(x)q(\theta|x)$ is updated to the
posterior $P(x,\theta)=P(x)P(\theta|x)$ where $P(x)=\delta(x-x^{\prime})$ is
fixed by the observed data while $P(\theta|x^{\prime})=q(\theta|x^{\prime})$
remains unchanged. Note that in accordance with the PMU philosophy that drives
the ME method \emph{one only updates those aspects of one's beliefs for which
corrective new evidence has been supplied}. In \cite{Caticha Giffin
2006}\cite{Caticha 2020}\cite{Giffin Caticha 2007} further examples are given
that show how ME allows generalizations of Bayes' rule to situations where the
data itself is uncertain, or there is information about moments of $x$, or
moments of $\theta$, or even in situations where the likelihood function is
unknown. In conclusion, the ME method of maximum entropy can fully reproduce
and then go beyond the results obtained by the standard Bayesian methods.

\section{Deviations from maximum entropy}

The basic ME problem is to update from a prior $q(x)$ given information
specified by certain constraints. The constraints specify a family of
candidate distributions
\begin{equation}
p_{\theta}(x)=p(x|\theta) \label{ptheta}%
\end{equation}
which can be conveniently labelled with a finite number of parameters
$\theta^{a}$, $a=1\ldots n$. (The generalization to an infinite number of
parameters poses technical but not insurmountable difficulties.) Thus, the
parameters $\theta$ are coordinates on the statistical manifold specified by
the constraints. The distributions in this manifold are ranked according to
their entropy
\begin{equation}
S[p_{\theta},q]=-\int dx\,p(x|\theta)\,\log\frac{p(x|\theta)}{q(x)}=S(\theta)
\label{S theta}%
\end{equation}
and the selected posterior is the distribution $p(x|\theta_{0})$ that
maximizes the entropy $S(\theta)$. (The notation indicates that $S[p_{\theta
},q]$ is a functional of $p_{\theta}$ while $S(\theta)$ is a function of
$\theta$.)

The question we now address concerns the extent to which $p(x|\theta_{0})$
should be preferred over other distributions with lower entropy or, to put it
differently: To what extent is it rational to believe that the selected value
ought to be the entropy maximum $\theta_{0}$ rather than any other value
$\theta$? \cite{Caticha 2003} This is a question about the probability
$p(\theta)$ of various values of $\theta$. The original problem which led us
to design the maximum entropy method was to assign a probability to the
quantity $x$; we now see that the full problem is to assign probabilities to
both $x$ and $\theta$.\ We are concerned not just with $p(x)$ but rather with
the joint distributions which we denote $\pi(x,\theta)$; the universe of
discourse has been expanded from $\mathcal{X}$ (the space of $x$s) to the
product space $\mathcal{X}\times\Theta$ ($\Theta$ is the space of parameters
$\theta$).

To determine the joint distribution $\pi(x,\theta)$ we make use of essentially
the only (universal) method at our disposal --- the ME method itself --- but
this requires that we address the standard two preliminary questions: first,
what is the prior distribution? What do we know about $x$ and $\theta$ before
we receive information about the constraints? And second, what is the new
information that constrains the allowed joint distributions $\pi(x,\theta)$?

This first question is the more subtle one: when we know absolutely nothing
about the $\theta$s we know neither their physical meaning nor whether there
is any relation to the $x$s. A joint prior that reflects this lack of
correlations is a product, $q(x,\theta)=q(x)q(\theta)$. We will assume that
the prior $q(x)$ is known --- it is the same prior we had used when we updated
from $q(x)$ to $p(x|\theta_{0})$ using (\ref{S theta}).

But we are not totally ignorant about the $\theta$s: we know that they label
distributions $\pi(x|\theta)$ on some as yet unspecified statistical manifold
$\Theta$. Then there exists a natural measure of distance in the space
$\Theta$. It is given by the information metric $d\ell^{2}=g_{ab}d\theta
^{a}d\theta^{b}$ \cite{Caticha 2020}\cite{Amari 1985},
\begin{equation}
g_{ab}=\int dx\,p(x|\theta)\frac{\partial\log p(x|\theta)}{\partial\theta^{a}%
}\frac{\partial\log p(x|\theta)}{\partial\theta^{b}}. \label{Fisher metric}%
\end{equation}
and the corresponding volume elements are given by $g^{1/2}(\theta)d^{n}%
\theta$, where $g(\theta)$ is the determinant of the metric. The uniform prior
for $\theta$, which assigns equal probabilities to equal volumes, is
proportional to $g^{1/2}(\theta)$ and therefore we choose $q(\theta
)=g^{1/2}(\theta)$. Therefore, the joint prior is $q(x,\theta)=q(x)g^{1/2}%
(\theta)$.

Next we tackle the second question: what are the constraints on the allowed
joint distributions $\pi(x,\theta)$? Consider the space of all joint
distributions. To each choice of the functional form of $\pi(x|\theta)$ (for
example, whether we talk about Gaussians, Boltzmann-Gibbs distributions, or
something else) there corresponds a different subspace defined by
distributions of the form $\pi(x,\theta)=\pi(\theta)\pi(x|\theta)$. The
crucial constraint is that which specifies the subspace by imposing that
$\pi(x|\theta)$ take the particular functional form given by the constraint
(\ref{ptheta}), $\pi(x|\theta)=p(x|\theta)$. This defines the meaning to the
$\theta$s and also fixes the prior $g^{1/2}(\theta)$ on the relevant subspace.

The preferred joint distribution, $P(x,\theta)=P(\theta)p(x|\theta)$, is the
distribution, $\pi(x,\theta)=\pi(\theta)p(x|\theta)$, that maximizes the joint
entropy,
\begin{align}
\mathcal{S}[\pi,q]  &  =-\int dx\,d\theta\,\pi(\theta)p(x|\theta)\,\log
\frac{\pi(\theta)p(x|\theta)}{g^{1/2}(\theta)q(x)}\nonumber\\
&  =-\int\,d\theta\,\pi(\theta)\log\frac{\pi(\theta)}{g^{1/2}(\theta)}+\int
d\theta\,\pi(\theta)S(\theta)~, \label{S[joint]}%
\end{align}
where $S(\theta)$ is given in (\ref{S theta}). Varying (\ref{S[joint]}) with
respect to $\pi(\theta)$ with $\int d\theta\,\pi(\theta)=1$ and $p(x|\theta)$
fixed yields the posterior probability that the value of $\theta$ lies within
the small volume $g^{1/2}(\theta)d^{n}\theta$,
\begin{equation}
P(\theta)d^{n}\theta=\frac{1}{\zeta}\,\,e^{S(\theta)}g^{1/2}(\theta
)d^{n}\theta\quad\text{with\quad}\zeta=\int d^{n}\theta\,g^{1/2}%
(\theta)\,e^{S(\theta)}. \label{main}%
\end{equation}
Equation (\ref{main}) is the result we seek. It tells us that, as expected,
the preferred value of $\theta$ is the value $\theta_{0}$ that maximizes the
entropy $S(\theta)$, eq.(\ref{S theta}), because this maximizes the scalar
density $\exp S(\theta)$. But it also tells us the degree to which values of
$\theta$ away from the maximum are ruled out. (Note that the density $\exp
S(\theta)$ is a scalar function and the presence of the Jacobian factor
$g^{1/2}(\theta)$ makes eq.(\ref{main}) manifestly invariant under changes of
the coordinates $\theta$ in the space $\Theta$.)

This discussion allows us to refine our understanding of the ME method. ME\ is
not an all-or-nothing recommendation to pick the single distribution that
maximizes entropy and reject all others. The ME\ method is more nuanced: in
principle all distributions within the constraint manifold ought to be
included in the analysis; they contribute in proportion to the exponential of
their entropy and this turns out to be significant in situations where the
entropy maximum is not particularly sharp.

Going back to the original problem of updating from the prior $q(x)$ given
information that specifies the manifold $\{p(x|\theta)\}$, the preferred
update within the family $\{p(x|\theta)\}$ is $p(x|\theta_{0})$, but to the
extent that other values of $\theta$ are not totally ruled out, a better
update is obtained marginalizing the joint posterior $P(x,\theta
)=P(\theta)p(x|\theta)$ over $\theta$,
\begin{equation}
P(x)=\int d^{n}\theta\,P(\theta)p(x|\theta)=\int d^{n}\theta\,g^{1/2}%
(\theta)\frac{\,e^{S(\theta)}}{\zeta}p(x|\theta)~. \label{ME posterior}%
\end{equation}
In situations where the entropy maximum at $\theta_{0}$ is very sharp we
recover the old result,
\begin{equation}
P(x)\approx p(x|\theta_{0})~.
\end{equation}
When the entropy maximum is not very sharp a more honest update is
eq.(\ref{ME posterior}) which, incidentally, is a form of superstatistics.

One of the limitations of the standard MaxEnt method is that it selects a
single \textquotedblleft posterior\textquotedblright\ $p(x|\theta_{0})$ and
strictly rules out all other distributions. The result (\ref{main}) overcomes
this limitation and finds many applications. For example, it extends the
Einstein theory of thermodynamic fluctuations beyond the regime of small
fluctuations; it provides a bridge to the theory of large deviations; and,
suitably adapted for Bayesian data analysis, it leads to the notion of
entropic priors \cite{Caticha Preuss 2004}.

\section{Discussion}

\paragraph*{\noindent Consistency with the law of large numbers ---}

Entropic methods of inference are of general applicability but there exist
special situations --- such as, for example, those involving large numbers of
independent subsystems --- where inferences can be made by purely
probabilistic methods without ever invoking the concept of entropy. In such
cases one can check (see \emph{e.g. }in \cite{Caticha 2008}\cite{Grendar
2003}) that the two methods of calculation are consistent with each other. It
is significant, however, that alternative entropies such as those proposed in
\cite{Renyi 1961}\cite{Aczel Daroczy 1975}\cite{Tsallis 1988} do not pass this
test \cite{LaCour Schieve 2000}\cite{Nauenberg 2003}, which rules them out as
tools for updating. Some probability distributions obtained by maximizing the
alternative entropies have, however, turned out to be physically relevant. It
is therefore noteworthy that those successful distributions can also be
derived through more standard application of MaxEnt or ME as advocated in this
review \cite{Caticha 2020}\cite{Plastino Plastino 1994}\cite{Beck
2001}\cite{Beck Cohen 2003}\cite{Crooks 2007}. In other words, what is being
ruled out are not the distributions themselves, but the alternative entropies
from which they were inferred.

\paragraph*{\noindent On priors ---}

Choosing the prior density $q(x)$ can be tricky. Sometimes symmetry
considerations can be useful but otherwise there is no fixed set of rules to
translate information into a probability distribution except, of course, for
Bayes' rule and the ME method themselves.

What if the prior $q(x)$ vanishes for some values of $x$? $S[p,q]$ can be
infinitely negative when $q(x)$ vanishes within some region $\mathcal{D}$.
This means that the ME method confers an infinite preference on those
distributions $p(x)$ that vanish whenever $q(x)$ does. One must emphasize that
this is as it should be. A similar situation\ also arises in the context of
Bayes' theorem where assigning a vanishing prior represents a tremendously
serious commitment because no amount of data to the contrary would allow us to
revise it. In both ME and Bayes updating we should recognize the implications
of assigning a vanishing prior. Assigning a very low but non-zero prior
represents a safer and possibly less prejudiced representation of one's prior beliefs.

\paragraph*{Commuting and non-commuting constraints ---}

The ME method allows one to process information in the form of constraints.
When we are confronted with several constraints we must be particularly
cautious. Should they be processed simultaneously or sequentially? And, if the
latter, in what order? The answer depends on the problem at hand \cite{Giffin
Caticha 2007}.

We refer to constraints as \emph{commuting} when it makes no difference
whether they are handled simultaneously or sequentially. The most common
example is that of Bayesian updating on the basis of data collected in several
independent experiments. In this case the order in which the observed data
$x^{\prime}=\{x_{1}^{\prime},x_{2}^{\prime},\ldots\}$ is processed does not
matter for the purpose of inferring $\theta$. In general, however, constraints
need not commute and when this is the case the order in which they are
processed is critical.

To decide whether constraints are to be handled sequentially or simultaneously
one must be clear about how the ME method handles constraints. The ME
machinery interprets a constraint in a very mechanical way: all distributions
satisfying the constraint are in principle allowed while all distributions
violating it are ruled out. Therefore, sequential updating is appropriate when
old constraints become obsolete and are superseded by new information while
simultaneous updating is appropriate when old constraints remain valid. The
two cases refer to different states of information and therefore it is to be
expected that they will result in different inferences. These comments are
meant to underscore the importance of understanding what information is and
how it is processed by the ME method; failure to do so will lead to errors
that do not reflect a shortcoming of the ME method but rather a misapplication
of it.

\paragraph*{Pitfalls? ---}

Entropy is a tool for reasoning and --- as with all tools for reasoning or
otherwise --- it can be misused leading to unsatisfactory results
\cite{Caticha 2012b}. Should that happen, the inevitable questions are `what
went wrong?' and `how do we fix it?' It helps to first ask what components of
the analysis can be trusted so that the possible mistakes can be looked for
elsewhere. The answers proposed by the ME method are radically conservative:
problems always arise through a wrong choices of variables, or priors, or
constraints. Indeed, one should not blame the entropic method for not having
discovered and taken into account relevant information that was not explicitly
introduced into the analysis. Indeed, just as one would be very reticent about
questioning the basic rules of arithmetic, or the basic rules of calculus, one
should not question the basic sum and product rules of the probability
calculus and, taking this one step farther, one should not question the
applicability of entropy as the updating tool. The adoption of this
conservative approach leads us to reject alternative entropies and quantum
probabilities. Fortunately, those constructs are not actually needed --- as
mentioned above, those Tsallis distributions that have turned out be useful
can be derived with the standard entropic methods \cite{Caticha 2020}%
\cite{Plastino Plastino 1994}\cite{Beck 2001}\cite{Beck Cohen 2003}%
\cite{Crooks 2007} and quantum mechanics can be handled within standard
probability theory without invoking exotic probabilities \cite{Caticha
2019}\cite{Bohm Hiley 1993}.

\paragraph{Acknowledgments ---}

I would like to acknowledge many valuable discussions on probability and
entropy with N. Caticha, A. Giffin, K. Knuth, R. Preuss, C. Rodr\'{\i}guez, J.
Skilling, and K. Vanslette.

\section*{Appendix A: DC1 -- Mutually exclusive subdomains}

In these appendices we establish the consequences of the two criteria DC1 and
DC2 leading to the final result eq.(\ref{S[p,q]}). The details of the proofs
are important not just because they lead to our final conclusions, but also
because the translation of the verbal statement of the criteria into precise
mathematical form is a crucial part of unambiguously specifying what the
criteria actually say.

First we prove that criterion DC1 leads to the expression eq.(\ref{DC1b}) for
$S[p,q]$. Consider the case of a discrete variable, $p_{i}$ with $i=1\ldots
n$, so that $S[p,q]=S(p_{1}\ldots p_{n},q_{1}\ldots q_{n})$. Suppose the space
of states $\mathcal{X}$ is partitioned into two non-overlapping domains
$\mathcal{D}$ and $\mathcal{\tilde{D}}$ with $\mathcal{D\cup\tilde{D}%
}=\mathcal{X}$, and that the information to be processed is in the form of a
constraint that refers to the domain $\mathcal{\tilde{D}}$,
\begin{equation}%
{\textstyle\sum\limits_{j\in\mathcal{\tilde{D}}}}
a_{j}p_{j}=A\text{ .} \tag{A1}\label{locality a}%
\end{equation}
DC1 states that the constraint on $\mathcal{\tilde{D}}$ does not have an
influence on the \emph{conditional} probabilities $p_{i|\mathcal{D}}$. It may
however influence the probabilities $p_{i}$ within $\mathcal{D}$ through an
overall multiplicative factor. To deal with this complication consider then a
special case where the overall probabilities of $\mathcal{D}$ and
$\mathcal{\tilde{D}}$ are constrained too,
\begin{equation}%
{\textstyle\sum\limits_{i\in\mathcal{D}}}
p_{i}=P_{\mathcal{D}}\quad\text{and \quad}%
{\textstyle\sum\limits_{j\in\mathcal{\tilde{D}}}}
p_{j}=P_{\mathcal{\tilde{D}}}\text{ }, \tag{A2}\label{locality b}%
\end{equation}
with $P_{\mathcal{D}}+P_{\mathcal{\tilde{D}}}=1$. Under these special
circumstances constraints on $\mathcal{\tilde{D}}$ will not influence $p_{i}$s
within $\mathcal{D}$, and vice versa.

To obtain the posterior maximize $S[p,q]$ subject to these three constraints,
\begin{align*}
0  &  =\left[  \delta S-\lambda\left(
{\textstyle\sum\limits_{i\in\mathcal{D}}}
p_{i}-P_{\mathcal{D}}\right)  +\right. \\
&  -\left.  \tilde{\lambda}\left(
{\textstyle\sum\limits_{j\in\mathcal{\tilde{D}}}}
p_{i}-P_{\mathcal{\tilde{D}}}\right)  +\mu\left(
{\textstyle\sum\limits_{j\in\mathcal{\tilde{D}}}}
a_{j}p_{j}-A\right)  \right]  ~,
\end{align*}
leading to
\begin{align}
\frac{\partial S}{\partial p_{i}}  &  =\lambda\text{\quad for\quad}%
i\in\mathcal{D}~,\tag{A3}\label{locality c}\\
\frac{\partial S}{\partial p_{j}}  &  =\tilde{\lambda}+\mu a_{j}\text{\quad
for\quad}j\in\mathcal{\tilde{D}}~. \tag{A4}\label{locality d}%
\end{align}
Eqs.(\ref{locality a}-\ref{locality d}) are $n+3$ equations we must solve for
the $p_{i}$s and the three Lagrange multipliers, $\lambda$, $\tilde{\lambda}$,
and $\mu$. Since $S=S(p_{1}\ldots p_{n},q_{1}\ldots q_{n})$ its derivative
\[
\frac{\partial S}{\partial p_{i}}=f_{i}(p_{1}\ldots p_{n},q_{1}\ldots q_{n})
\]
could in principle also depend on all $2n$ variables. But this violates the
DC1 criterion because any arbitrary change in $a_{j}$ within $\mathcal{\tilde
{D}}$ would influence the $p_{i}$s within $\mathcal{D}$. The only way that
probabilities conditioned on $\mathcal{D}$ can be shielded from arbitrary
changes in the constraints pertaining to $\mathcal{\tilde{D}}$ is that for any
$i\in\mathcal{D}$ the function $f_{i}$ depends only on $p_{j}$s with
$j\in\mathcal{D}$. Furthermore, this must hold not just for one particular
partition of $\mathcal{X}$ into domains $\mathcal{D}$ and $\mathcal{\tilde{D}%
}$, it must hold for \emph{all conceivable partitions} including the partition
into atomic propositions. Therefore $f_{i}$ can depend only on $p_{i}$,
\begin{equation}
\frac{\partial S}{\partial p_{i}}=f_{i}(p_{i},q_{1}\ldots q_{n})~.
\tag{A5}\label{locality e}%
\end{equation}

But the power of the criterion DC1 is not exhausted yet. The information that
affects the posterior can enter not just through constraints but also through
the prior. Suppose that the local information about domain $\mathcal{\tilde
{D}}$ is altered by changing the prior within $\mathcal{\tilde{D}}$. Let
$q_{j}\rightarrow q_{j}+\delta q_{j}$ for $j\in\mathcal{\tilde{D}}$. Then
(\ref{locality e}) becomes
\[
\frac{\partial S}{\partial p_{i}}=f_{i}(p_{i},q_{1}\ldots q_{j}+\delta
q_{j}\ldots q_{n})
\]
which shows that $p_{i}$ with $i\in\mathcal{D}$ will be influenced by
information about $\mathcal{\tilde{D}}$ unless $f_{i}$ with $i\in\mathcal{D}$
is independent of all the $q_{j}$s for $j\in\mathcal{\tilde{D}}$. Again, this
must hold for all possible partitions into $\mathcal{D}$ and $\mathcal{\tilde
{D}}$, and therefore,
\[
\frac{\partial S}{\partial p_{i}}=f_{i}(p_{i},q_{i})\text{\quad for all\quad
}i\in\mathcal{X}~.
\]

The choice of the functions $f_{i}(p_{i},q_{i})$ can be restricted further. If
we were to maximize $S[p,q]$ subject to constraints
\[%
{\textstyle\sum\nolimits_{i}}
p_{i}=1\quad\text{and}\quad%
{\textstyle\sum\nolimits_{i}}
a_{i}p_{i}=A
\]
we get
\[
\frac{\partial S}{\partial p_{i}}=f_{i}(p_{i},q_{i})=\lambda+\mu
a_{i}\text{\quad for all\quad}i\in\mathcal{X}~,
\]
where $\lambda$ and $\mu$ are Lagrange multipliers. Solving for $p_{i}$ gives
a posterior,
\[
P_{i}=g_{i}(q_{i},\lambda,\mu,a_{i})
\]
for some functions $g_{i}$. As stated in Section \ref{specific DC} we do not
assume that the labels $i$ themselves carry any particular significance. This
means, in particular, that for any proposition labelled $i$ we want the
selected posterior $P_{i}$ to depend only on the numbers $q_{i}$, $\lambda$,
$\mu$, and $a_{i}$. We do not want to have different updating rules for
different propositions: two different propositions $i$ and $i^{\prime}$ with
the same $q_{i}=q_{i^{\prime}}$ and the same $a_{i}=a_{i^{\prime}}$ should be
updated to the same posteriors, $P_{i}=P_{i^{\prime}}$. In other words, the
functions $g_{i}$ and $f_{i}$ must be independent of $i$. Therefore
\begin{equation}
\frac{\partial S}{\partial p_{i}}=f(p_{i},q_{i})\text{\quad for all\quad}%
i\in\mathcal{X}~. \tag{A6}\label{dS/dp}%
\end{equation}
Integrating, one obtains%
\[
S[p,q]=%
{\textstyle\sum\nolimits_{i}}
F(p_{i},q_{i})+\operatorname{constant}\text{.}%
\]
for some still undetermined function $F$. The constant has no effect on the
entropy maximization and can be dropped.

The corresponding expression for a continuous variable $x$ is obtained
replacing $i$ by $x$, and the sum over $i$ by an integral over $x$ leading to
eq.(\ref{DC1b}),
\[
S[p,q]=\int dx\,F\left(  p(x),q(x)\right)  \ .
\]

\section*{Appendix B: DC2 -- Independent subsystems}

Here we show that DC2 leads to eq.(\ref{S[p,q]}). Let the microstates of a
composite system be labeled by $(i_{1},i_{2})\in\mathcal{X}=\mathcal{X}%
_{1}\times\mathcal{X}_{2}$. We shall consider two special cases.

\paragraph{Case (a) ---}

First we treat the two subsystems separately. Suppose that for subsystem 1 we
have the extremely constraining information that updates $q_{1}(i_{1})$ to be
$P_{1}(i_{1})$, and for subsystem 2 we have no new information at all. For
subsystem $1$ we maximize $S_{1}[p_{1},q_{1}]$ subject to the constraint
$p_{1}(i_{1})=P_{1}(i_{1})$ and the selected posterior is, of course,
$p_{1}(i_{1})=P_{1}(i_{1})$. For subsystem 2 we maximize $S_{2}[p_{2},q_{2}]$
subject only to normalization and there is no update, $P_{2}(i_{2}%
)=q_{2}(i_{2})$.

When the systems are treated jointly, however, the inference is not nearly as
trivial. We want to maximize the entropy of the joint system,
\[
S[p,q]=%
{\textstyle\sum\limits_{i_{1},i_{2}}}
F\left(  p(i_{1},i_{2}),q_{1}(i_{1})q_{2}(i_{2})\right)  ~,
\]
subject to the constraint on subsystem 1,
\[%
{\textstyle\sum\nolimits_{i_{1}}}
\,p(i_{1},i_{2})=P_{1}(i_{1})~.
\]
Notice that this is not just one constraint: we have one constraint for each
value of $i_{1}$, and each constraint must be supplied with its own Lagrange
multiplier, $\lambda_{1}(i_{1})$. Then,
\[
\delta\left[  S-%
{\textstyle\sum\nolimits_{i_{1}}}
\lambda_{1}(i_{1})\left(
{\textstyle\sum\nolimits_{i_{2}}}
\,p(i_{1},i_{2})-P_{1}(i_{1})\right)  \right]  =0~.
\]
The independent variations $\delta p(i_{1},i_{2})$ yield
\[
f\,\left(  p(i_{1},i_{2}),q_{1}(i_{1})q_{2}(i_{2})\right)  =\lambda_{1}%
(i_{1})~,
\]
where $f$ is given in (\ref{dS/dp}),
\[
\frac{\partial S}{\partial p}=\frac{\partial}{\partial p}F\left(  p,q_{1}%
q_{2}\right)  =f\left(  p,q_{1}q_{2}\right)  ~.
\]
Next we impose that the selected posterior is the product $P_{1}(i_{1}%
)q_{2}(i_{2})$. The function $f$ must be such that
\[
f\,\left(  P_{1}q_{2},q_{1}q_{2}\right)  =\lambda_{1}~.
\]
Since the RHS is independent of the argument $i_{2}$, the $f$ function must be
such that the $i_{2}$-dependence cancels out and this cancellation must occur
for all values of $i_{2}$ and all choices of the prior $q_{2}$. Therefore we
impose that for any value of $x$ the function $f(p,q)$ must satisfy
\[
f(px,qx)=f(p,q)~.
\]
Choosing $x=1/q$ we get
\begin{equation}
f\left(  \frac{p}{q},1\right)  =f(p,q)\quad\text{or}\quad\frac{\partial
F}{\partial p}=f(p,q)=\phi\left(  \frac{p}{q}\right)  ~. \tag{B1}%
\label{dS/dp b}%
\end{equation}
Thus, the function $f(p,q)$ has been reduced to a function $\phi(p/q)$ of a
single argument.

\paragraph{Case (b) ---}

Next we consider a situation in which both subsystems are updated by extremely
constraining information: when the subsystems are treated separately
$q_{1}(i_{1})$ is updated to $P_{1}(i_{1})$ and $q_{2}(i_{2})$ is updated to
$P_{2}(i_{2})$. When the systems are treated jointly we require that the joint
prior for the combined system $q_{1}(i_{1})q_{2}(i_{2})$ be updated to
$P_{1}(i_{1})P_{2}(i_{2})$.

First we treat the subsystems separately. Maximize the entropy of subsystem
1,
\[
S[p_{1},q_{1}]=%
{\textstyle\sum\nolimits_{i_{1}}}
F\,\left(  p_{1}(i_{1}),q_{1}(i_{1})\right)  \quad\text{subject to}\quad
p_{1}(i_{1})=P_{1}(i_{1})~.
\]
To each constraint --- one constraint for each value of $i_{1}$ --- we must
supply one Lagrange multiplier, $\lambda_{1}(i_{1})$. Then,
\[
\delta\left[  S-%
{\textstyle\sum\nolimits_{i_{1}}}
\lambda_{1}(i_{1})\left(  \,p(i_{1})-P_{1}(i_{1})\right)  \right]  =0~.
\]
Using eq.(\ref{dS/dp b}),
\[
\frac{\partial S}{\partial p_{1}}=\frac{\partial}{\partial p_{1}}F\left(
p_{1},q_{1}\right)  =\phi\left(  \frac{p_{1}}{q_{1}}\right)  ~,
\]
and, imposing that the selected posterior be $P_{1}(i_{1})$, we find that the
function $\phi$ must obey
\begin{equation}
\phi\left(  \frac{P_{1}(i_{1})}{q_{1}(i_{1})}\right)  =\lambda_{1}(i_{1})~.
\tag{B2}\label{phi dc2 a}%
\end{equation}
Similarly, for system 2 we find
\begin{equation}
\phi\left(  \frac{P_{2}(i_{2})}{q_{2}(i_{2})}\right)  =\lambda_{2}(i_{2})~.
\tag{B3}\label{phi dc2 b}%
\end{equation}

Next we treat the two subsystems jointly. Maximize the entropy of the joint
system,
\[
S[p,q]=%
{\textstyle\sum\limits_{i_{1},i_{2}}}
F\,\left(  p(i_{1},i_{2}),q_{1}(i_{1})q_{2}(i_{2})\right)  ~,
\]
subject to the following constraints on the joint distribution $p(i_{1}%
,i_{2})$:
\[%
{\textstyle\sum\nolimits_{i_{2}}}
\,p(i_{1},i_{2})=P_{1}(i_{1})\qquad\text{and\qquad}%
{\textstyle\sum\nolimits_{i_{1}}}
\,p(i_{1},i_{2})=P_{2}(i_{2})~.
\]
Again, there is one constraint for each value of $i_{1}$ and of $i_{2}$ and we
introduce Lagrange multipliers, $\eta_{1}(i_{1})$ or $\eta_{2}(i_{2})$. Then,
\[
\delta\left[  S-%
{\textstyle\sum\nolimits_{i_{1}}}
\eta_{1}(i_{1})\left(
{\textstyle\sum\nolimits_{i_{2}}}
\,p(i_{1},i_{2})-P_{1}(i_{1})\right)  -\{1\leftrightarrow2\}\right]  =0,
\]
where $\{1\leftrightarrow2\}$ indicates a third term, similar to the second,
with $1$ and $2$ interchanged. The independent variations $\delta
p(i_{1},i_{2})$ yield
\[
\phi\left(  \frac{p(i_{1},i_{2})}{q_{1}(i_{1})q_{2}(i_{2})}\right)  =\eta
_{1}(i_{1})+\eta_{2}(i_{2})~,
\]
and we impose that the selected posterior be the product $P_{1}(i_{1}%
)P_{2}(i_{2})$. Therefore, the function $\phi$ must be such that
\[
\phi\left(  \frac{P_{1}P_{2}}{q_{1}q_{2}}\right)  =\eta_{1}+\eta_{2}~.
\]
To solve this equation we take the exponential of both sides, let $\xi
=\exp\phi$, and rewrite as
\begin{equation}
\xi\left(  \frac{P_{1}P_{2}}{q_{1}q_{2}}\right)  e^{-\eta_{2}(i_{2})}%
=e^{\eta_{1}(i_{1})}~. \tag{B4}\label{csi a}%
\end{equation}
This shows that for any value of $i_{1}$, the dependences of the LHS on
$i_{2}$ through $P_{2}/q_{2}$ and $\eta_{2}$ must cancel each other out. In
particular, if for some subset of $i_{2}$s the subsystem 2 is updated so that
$P_{2}=q_{2}$, which amounts to no update at all, the $i_{2}$ dependence on
the left is eliminated but the $i_{1}$ dependence remains unaffected,
\[
\xi\left(  \frac{P_{1}}{q_{1}}\right)  e^{-\eta_{2}^{\prime}}=e^{\eta
_{1}(i_{1})}~.
\]
where $\eta_{2}^{\prime}$ is some constant independent of $i_{2}$. A similar
argument with $\{1\leftrightarrow2\}$ yields
\[
\xi\left(  \frac{P_{2}}{q_{2}}\right)  e^{-\eta_{1}^{\prime}}=e^{\eta
_{2}(i_{2})}~,~
\]
where $\eta_{1}^{\prime}$ is a constant. Taking the exponential of
(\ref{phi dc2 a}) and (\ref{phi dc2 b}) leads to
\[
\xi\left(  \frac{P_{1}}{q_{1}}\right)  e^{-\eta_{2}^{\prime}}=e^{\lambda
_{1}-\eta_{2}^{\prime}}=e^{\eta_{1}}\quad\text{and}\quad\xi\left(  \frac
{P_{2}}{q_{2}}\right)  e^{-\eta_{1}^{\prime}}=e^{\lambda_{2}-\eta_{1}^{\prime
}}=e^{\eta_{2}}~.
\]
Substituting back into (\ref{csi a}), we get
\[
\xi\left(  \frac{P_{1}P_{2}}{q_{1}q_{2}}\right)  =\xi\left(  \frac{P_{1}%
}{q_{1}}\right)  \xi\left(  \frac{P_{2}}{q_{2}}\right)  ~,
\]
where a constant factor $e^{-(\eta_{1}^{\prime}+\eta_{2}^{\prime})}$ has been
absorbed into a new function $\xi$. The general solution of this functional
equation is a power,
\[
\xi(xy)=\xi(x)\xi(y)\Longrightarrow\xi(x)=x^{a}~,
\]
so that
\[
\phi(x)=a\log x+b~,
\]
where $a$ and $b$ are constants. Finally, integrate (\ref{dS/dp b}),
\[
\frac{\partial F}{\partial p}=\phi\left(  \frac{p}{q}\right)  =a\log\frac
{p}{q}+b~,
\]
to get
\[
F[p,q]=ap\log\frac{p}{q}+b^{\prime}p+c
\]
where $b^{\prime}$ and $c$ are constants.

At this point the entropy takes the general form
\[
S[p,q]=%
{\textstyle\sum\nolimits_{i}}
\left(  ap_{i}\log\frac{p_{i}}{q_{i}}+b^{\prime}p_{i}+c\right)  .
\]
The additive constant $c$ may be dropped: it contributes a term that does not
depend on the probabilities and has no effect on the ranking scheme.
Furthermore, since $S[p,q]$ will be maximized subject to constraints that
include normalization the $b^{\prime}$ term has no effect on the selected
distribution and can be dropped too. Finally, the multiplicative constant $a$
has no effect on the overall ranking except in the trivial sense that
inverting the sign of $a$ will transform the maximization problem to a
minimization problem or vice versa. We can therefore set $a=-1$ so that
maximum $S$ corresponds to maximum preference, which gives us eq.(\ref{S[p,q]}%
) and concludes our derivation.

\end{document}